\documentclass[mathleft
]{an}
\usepackage{graphicx}
\usepackage{amssymb,amsmath}
\usepackage{times}
\begin{document}

\Pagespan{1}{4}
\Yearpublication{2012}%
\Yearsubmission{2012}%
\Month{}%
\Volume{}%
\Issue{}%

\title{X-ray observations of the merging cluster CIZA J2242.8+5301}

\author{G.~Ogrean\inst{1,2}\fnmsep\thanks{Corresponding author:
  \email{gogrean@hs.uni-hamburg.de}\newline}
\and M.~Br\"uggen\inst{1,2}
\and A.~Simionescu\inst{3}
\and H.~R\"ottgering\inst{4}
\and R.~J.~van~Weeren\inst{4,5}
\and J.~H.~Croston\inst{6}
\and M.~Hoeft\inst{7}
}
\titlerunning{X-ray observations of the merging cluster CIZA J2242.8+5301}
\authorrunning{G.~Ogrean et al.}
\institute{
Jacobs University Bremen, Campus Ring 1, 28759, Bremen, Germany
\and 
Hamburger Sternwarte, Gojenbergsweg 112, 21029, Hamburg, Germany
\and 
KIPAC, Stanford University, 452 Lomita Mall, Stanford, CA 94305, USA
\and 
Leiden Observatory, Leiden University, 2300 RA, Leiden, The Netherlands
\and
Netherlands Institute for Radio Astronomy (ASTRON), Postbus 2, 7990 AA
Dwingeloo, The Netherlands
\and
University of Southampton, Highfield, SO17 1BJ, Southampton, U.K.
\and
Thüringer Landessternwarte Tautenburg, Sternwarte 5, 07778, Tautenburg, Germany
}

\received{xx xxx 2012}
\accepted{xx xxx 2012}
\publonline{later}


\abstract{Multiwavelength studies of radio relics at merger shocks set powerful constraints on the relics origin and formation mechanism. However, for X-ray observations, a main difficulty is represented by the low X-ray surface brightness far out in the cluster outskirts, where relics are typically found. Here, we present \emph{XMM-Newton} results from a 130-ks observation of CIZA J2242.8+5301, a cluster at $z=0.19$ that hosts a double radio relic. We focus on the well-defined northern relic. There is a difference of $\sim 55\%$ between the temperature we measure behind the relic, and the temperature measured with \emph{Suzaku}. We analyse the reasons for this large discrepancy, and discuss the possibility of reliably measuring the temperature beyond the northern relic.
}

\maketitle

\section{Introduction}
\label{s:introduction}

Radio relics are diffuse, Mpc-scaled, arc-shaped sources, typically found at large cluster-centric radii. They are strongly polarized, with magnetic fields roughly aligned with the length of the relic (for reviews, see e.g., Br\"uggen et al. 2012, Feretti et al. 2012). The commonly-accepted theory explaining the origin of relics is that they trace cluster-merger shocks. The shocks will accelerate electrons and ions to relativistic energies, producing a population of synchrotron-emitting particles observable at radio frequencies. The shocks will also heat and compress the ICM thermal plasma, and these effects should be observable at X-ray wavelengths in the form of temperature and density jumps. The Mach number of a shock can be calculated from the spectral index at the front of the relic, assuming a certain acceleration mechanism and pre-existing particle population. At the same time, it can be calculated from the X-ray properties of the gas on both sides of a relic, using the Rankine-Hugoniot conditions. Therefore, combined radio and X-ray observations of relic-hosting galaxy cluster mergers allow us to shed light on the origin of radio relics by testing correlations between the thermal and non-thermal effects of cluster merger shocks.

So far, only two merger shocks have been X-ray-confirmed at radio relics (Abell 754, Macario et al. 2011; Abell 3667, Finoguenov et al. 2010), even though the number of known radio relics is much larger ($\sim 50$, Feretti 2012). The major difficulty of X-ray observations is that relics are often detected at large distances from the cluster centre (Vazza et al. 2012), where the ICM surface brightness is very low. Even if cluster emission is present at these distances, modelling the spectra is complicated by confusion with the background. Therefore, for correctly measuring the thermal properties of the gas and the shock Mach number, it is essential to accurately model the sky background, and to characterize the systematic errors introduced by uncertainties in the instrumental and sky backgrounds.

The sky background is composed of thermal foreground Galactic emission, which dominates at soft energies, and of the cosmic X-ray background (CXB; emission from unresolved sources), which dominates at high energies. The CXB is well-described by a power-law with a photon index of 1.41 (De Luca \& Molendi 2004). At high Galactic latitudes, $|b| \gtrsim 20^{\circ}$, the foreground is typically modelled as the sum of Local Hot Bubble emission (unabsorbed, $T_{\rm X}\sim 0.1$ keV) and Galactic Halo emission (absorbed, $T_{\rm X}\sim 0.2$ keV). Close to the Galactic plane, the hydrogen column density increases, and the foreground emission becomes more complex. Indeed, additional thermal or line components often need to be included in the foreground model. Masui et al. (2009) detected additional foreground emission from a \emph{Suzaku} observation in the Galactic direction ($255^{\circ}$, $0^{\circ}$), which is best-described by a thermal component with $T \sim 0.75$ keV. Simionescu et al. (2011), in their analysis of \emph{Suzaku} observations of the Perseus cluster ($b \approx 15^{\circ}$), found a thermal foreground component with $T \approx 0.6$ keV. George et al. (2009) measured densities and temperatures in PKS 0745-191 beyond the virial radius ($r_{\rm 200}$; the radius at which the mean enclosed density is 200 times the critical density of the Universe at the redshift of the cluster), but their results were challenged by Eckert et al. (2011), who discovered a discrepancy between the \emph{Suzaku} and \emph{ROSAT} surface brightness profiles, most likely attributed to an additional foreground component that was not taken into account by George et al. (2009). Later \emph{Suzaku} observations near PKS 0745-191 allowed for an improved background analysis (Walker et al. 2012), showing the need for an additional thermal foreground component with $T\approx 0.6$ keV; interestingly, the background fit was not improved by the inclusion of the Galactic Halo component. These examples serve as cautionary tales about the importance of accurate background modelling, in particular when analysing the faint outskirts of galaxy clusters.

At the redshift of CIZA J2242.8+5301, 1 arcmin corresponds to 192 kpc. All the errors are given at the $1\sigma$ level.

\section{CIZA J2242.8+5301}
\label{s:cizaj2242}

CIZA J2242.8+5301, a merging cluster at $z=0.19$, hosts a spectacular northern radio relic: 2-Mpc long, {$\sim75$-kpc} wide, strongly polarized ($\gtrsim 50\%$), and with magnetic fields aligned with the relic (van Weeren et al. 2010). The merger takes place in the plane of the sky (van Weeren et al. 2011), which reduces the number of unknowns that can be introduced by projection effects. Therefore, CIZA J2242.8+5301 is as a textbook example for studying the connection between radio relics and merger shocks. 

We have successfully proposed for \emph{XMM-Newton} and \emph{Chandra} observations of CIZA J2242.8+5301. The main results of these observations (130 and 200 ks, respectively) will be discussed in Ogrean et al. (2012) and Ogrean et al. (in preparation). CIZA J2242.8+5301 was also observed with \emph{Suzaku} (Akamatsu \& Kawahara 2011; hereafter, AK11). Worryingly, there appear to be significant differences between the \emph{XMM-Newton}- and \emph{Suzaku}-measured temperatures, which further affect the X-ray-derived shock Mach number at the northern relic. Here, we present the \emph{XMM-Newton} spectral analysis of the sky background in the direction of the cluster and the ICM properties across the northern relic, and discuss possible reasons for the discrepancy between the \emph{XMM-Newton} and \emph{Suzaku} results.

\begin{table*}
 \caption{Best-fit background parameters. Unstarred values were obtained using a background model that includes HF emission. Starred values were obtained with a background model without HF emission, as used by AK11 (see Section \ref{s:discussion}). Temperatures are in units of keV. Power-law normalizations are in units of photons\,keV$^{-1}$\,cm$^{-2}$\,s$^{-1}$\,arcmin$^{-2}$ at 1 keV. The normalizations of the APEC components are in units of cm$^{-5}$\,arcmin$^{-2}$. Values marked with a dagger ($\dagger$) were fixed in the fit.}
\begin{center}
 \footnotesize{
 \begin{tabular}{ccccccc}
  \hline
   Component & $\Gamma$ & $T_{\rm X}$ & $\mathcal{N}$ & $\Gamma^*$ & $T_{\rm X}^*$ & $\mathcal{N}^*$ \\
  \hline
    LHB & -- & $0.08^\dagger$ & $7.8_{-1.7}^{+1.4} \times 10^{-6}$ & -- & $0.08^\dagger$ & $1.0_{-0.062}^{+0.055} \times 10^{-5}$ \\
    GH  & -- & $0.20_{-0.019}^{+0.028}$ & $5.1_{-1.3}^{+1.9} \times 10^{-6}$ & -- & $0.28_{-0.012}^{+0.0083}$ & $3.4_{-0.18}^{+0.31} \times 10^{-6}$ \\
    HF  & -- & $0.79_{-0.058}^{+0.11}$ & $5.8_{-1.4}^{+1.2} \times 10^{-7}$ -- & -- & -- \\
    power-law & $1.41^\dagger$ & -- & $(1.01\pm 0.048) \times 10^{-6}$ & $1.41^\dagger$ & -- & $(1.12\pm 0.045) \times 10^{-6}$ \\
  \hline
 \end{tabular}}
\end{center}
 \label{tab:bestfitbkg}
\end{table*}

\section{Spectral analysis}
\label{s:spectralanalysis}

We reduced the data using the \emph{XMM-Newton} Extended Source Analysis Software ({\sc esas}), integrated in the Scientific Analysis System ({\sc sas}) v12.0.1. The details of the data reduction can be found in Ogrean et al. (2012).

\begin{figure}
\centering
\includegraphics[scale=0.35]{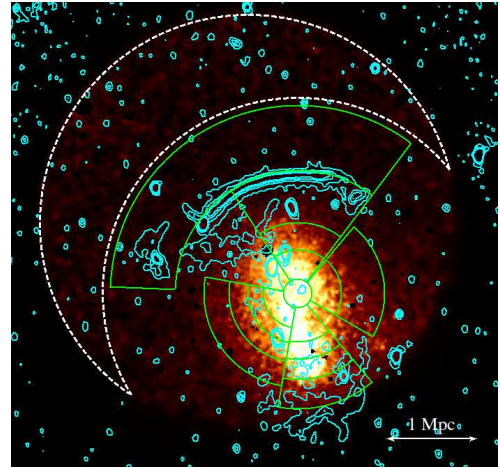}
\caption{\emph{XMM-Newton} MOS+pn $0.5-4$ keV exposure-corrected image, with the instrumental background subtracted. Overlaid in green are the regions used for extracting spectra. The dashed moon-shaped region shows the defined background. In cyan are \emph{WSRT} 1.4 GHz radio contours, drawn at $\left[1,5,10,...\right]\times 100$ $\mu$Jy/beam.}
\label{fig:regions}
\end{figure}

Spectra were extracted from the regions shown in Figure \ref{fig:regions}, which are all located within 12 arcmin of the detector centre. The background region is beyond $r_{\rm 200}\approx 2$ Mpc (Ogrean et al. 2012) from the merger centre, and should therefore contain only minimal cluster emission. For each region, we also created instrumental background spectra. Source and sky background spectra were grouped to a minimum of 30 counts per bin.

Modelling the background in the direction of the cluster is complicated by the cluster's low Galactic latitude ($b\sim-6^{\circ}$). The best fit to the background is given by a model consisting of CXB, LHB, GH, and hot foreground (HF; Masui et al. 2009, Simionescu et al. 2012) emission (see Section \ref{s:discussion}). We used {\sc apec} models for the thermal components (LHB, GH, HF), and a {\sc pow} model for the CXB component. Absorbed components (GH, HF) were described by {\sc wabs*apec} models, with coupled, but free, hydrogen column densities. The GH temperature was fixed to 0.08 keV, the average of the values found by Sidher et al. (1996) and Kuntz \& Snowden (2000). The hydrogen column density was frozen to $3.34\times 10^{21}$ cm$^{-2}$, corresponding to the 21-cm line measurement of the Galactic hydrogen column density (Dickey \& Lockman 1990) at the position of the cluster.

Cluster emission in all regions was modelled as a single-temperature, absorbed thermal component ({\sc wabs*apec}). The temperature and normalizations were free in the fit, but coupled for the MOS and pn spectra extracted from identical sky regions. The metallicity was fixed to 0.2 solar, and the cluster redshift to 0.1921 (Kocevski et al. 2007).

To maximize the number of constraints, the sky background and target spectra were fitted simultaneously. The instrumental background was substracted before fitting. The fits were performed in {\sc Xspec} v12.7.1, using Anders \& Grevesse (1989) abundances, and Balucinska-Church \& McCammon (1992) photoelectric absorption cross-sections. The fit had a $\chi^2$ of 1985.38, with 1833 degrees of freedom. The best-fit parameters are summarized in Tables \ref{tab:bestfitbkg} and \ref{tab:bestfitannuli}. The temperature beyond the relic is not constrained, hence we do not include it in Table \ref{tab:bestfitannuli}.


\begin{table}
 \caption{Best-fit temperatures for the partial annuli in the northern sector, with ($T_{\rm X}$) and without ($T_{\rm X}^*$) a HF component included in the background model (see Section \ref{s:discussion}).}
\begin{center}
 \footnotesize{
 \begin{tabular}{ccc}
  \hline
   $r$ (arcsec) & $T_{\rm X}$ (keV) & $T_{\rm X}^*$ (keV) \\
  \hline
    $0-50$ & $8.3_{-0.49}^{+0.58}$ & $8.2_{-0.49}^{+0.55}$ \\
    $50-150$ & $8.2_{-0.47}^{+0.55}$ & $8.1_{-0.47}^{+0.55}$ \\
    $150-245$ & $10.7_{-1.0}^{+1.2}$ & $10.2_{-1.0}^{+1.1}$ \\
    $245-425$ & $15.8_{-4.4}^{+6.1}$ & $10.8_{-2.0}^{+3.6}$  \\
    $425-650$ & -- & $1.7_{-0.34}^{+0.31}$ \\
  \hline
 \end{tabular}}
\end{center}
 \label{tab:bestfitannuli}
\end{table}

\section{Discussion and conclusions}
\label{s:discussion}

Unfortunately, the \emph{XMM-Newton} data does not allow us to constrain the temperature beyond the northern relic. The radio spectral index is $-0.6\pm 0.05$, which in the linear regime corresponds to a (putative) shock Mach number of $4.6_{-0.9}^{+1.3}$. In the X-ray, we do not see a temperature or density jump across the relic. However, assuming the relic traces a shock of Mach number $4.6$, a post-shock temperature of $\sim 15$~keV implies a pre-shock temperature of $\sim 2$~keV. AK11, using \emph{Suzaku} observations, measured a pre-shock temperature of $1.7\pm 0.20$~keV, although they do not detect a surface brightness jump across the relic; additionally, they measured a post-shock temperature of $6.7\pm 0.45$~keV, more than two times lower than the \emph{XMM-Newton} temperature measured in approximately the same region. In fact, the highest temperature in the \emph{Suzaku} temperature profile across the northern relic is $\sim 8$ keV. One major difference between our analysis and that of AK11 is the background modelling. Unlike us, they did not include a HF component. We modelled the \emph{XMM-Newton} spectra using the same model as AK11, to see if the different background model can account for the temperature discrepancies. For this, we simply needed to remove the HF component from the model described in Section \ref{s:spectralanalysis}. The results of the sky background and ICM spectral fits are summarized in Tables \ref{tab:bestfitbkg} and \ref{tab:bestfitannuli}.

An F-test was run to check whether it is truly reasonable to add a HF component to the background model. This yields a probability of $\sim 10^{-11}$. The low value of the probability indicates that, based on the \emph{XMM-Newton} data, the background is more accurately described by a model that includes a HF component.

When using the same background components as AK11, the \emph{XMM-Newton} temperatures are significantly lowered, especially at large distances from the merger's centre, and we also derive a pre-shock temperature that is consistent with that of AK11. However, the \emph{XMM-Newton} and \emph{Suzaku} results are still not consistent in the post-shock region. Another candidate for the temperature discrepancy could be residual soft proton (SP) contamination of the \emph{XMM-Newton} event files. We checked the SP-filtered event files for residual SP contamination using the {\sc fin\_over\_fout} routine (De Luca \& Molendi 2004), in the energy bands $8-12$ and $10-12$ keV for MOS, and $12-14$ keV for pn event files. The event files were diagnosed as not contaminated by SP.

The difference between the \emph{XMM-Newton} and \emph{Suzaku} temperatures could be blamed on calibration uncertainties. We note that at the time of the \emph{Suzaku} observation and when the paper of AK11 was published, there were some uncertainties in the calibration of XIS0 and XIS1 detectors, introduced by the rising contamination layer thickness of XIS0, and the change in the charge injection level of XIS1. Unfortunately, we are not able to tell if these uncertainties are enough to explain the temperature discrepancy.

\begin{figure}
\includegraphics[width=0.48\textwidth]{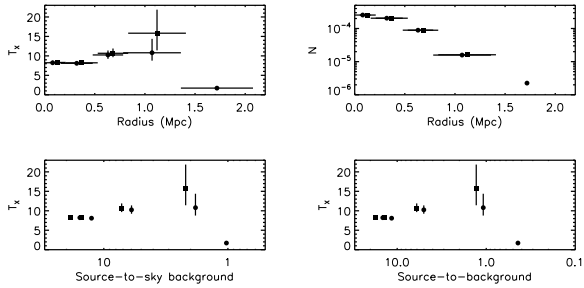}
\caption{\emph{Top:} Temperature and normalization profiles across the northern sector in Figure \ref{fig:regions}, using a background model with and without HF emission (squares and circles, respectively). \emph{Bottom:} Temperature profiles plotted against the source-to-sky background (left) and source-to-total background (right) count rate ratio. The profiles for the background model including HF emission are slightly shifted, for clarity.}
\label{fig:comparison}
\end{figure}

Another problem in measuring temperatures far out in the cluster outskirts is the systematic errors introduced by the background model. This could constitute an issue for measuring the pre-shock temperature at the northern relic. Leccardi \& Molendi (2008) analysed a sample of $\sim 50$ hot ($T>3.5$ keV) galaxy clusters at intermediate redshifts ($0.1\lesssim z \lesssim 0.3$) from the \emph{XMM-Newton} archive, and showed that for source-to-background count rate ratios below 0.6 in the energy band $0.7-10.0$ keV, the temperature can be biased downwards by a factor of a few. In Figure \ref{fig:comparison}, we show the temperature profile across the northern sector as a function of source-to-background count rate ratio in the energy band $0.5-7.0$ keV. For the background, we assumed the background region shown in Figure \ref{fig:regions}. The source-to-background count rate ratio is less than 0.5 in the pre-shock region. Compared to the threshold ratio set by Leccardi \& Molendi (2008), our ratio is probably even less encouraging, given that the threshold in the $0.5-7.0$ keV band should be higher than in $0.7-10.0$ keV (due to lower background). Therefore, it is not surprising that we are not able to properly constrain the pre-shock temperature from our \emph{XMM-Newton} observation. At the moment, no study similar to that of Leccardi \& Molendi (2008) was done for \emph{Suzaku}, so it is difficult to say how much the temperature measurement in the pre-shock region is affected by background systematics.


In conclusion:

1. When modelling the sky background at low Galactic latitudes, it is essential to consider all possible foreground components. In the case of CIZA J2242.8+5301, if the HF component is not included in the spectral model, the post-shock temperature is biased downwards by $\sim 30\%$. If the northern relic traces a merger shock, then the measurement of the colder pre-shock temperature will be less affected by the incorrect background model, as the HF and ICM temperatures are more similar in the pre-shock region; hence, this will also bias downwards the X-ray-derived Mach number.

2. Using the presented \emph{XMM-Newton} observation, it is impossible to measure the temperature beyond the northern relic in CIZA J2242.8+5301 while keeping systematic uncertainties under control. This might be possible with \emph{Suzaku}, due to its lower instrumental background. However, lacking statistical studies about the necessary source-to-background count rate threshold for \emph{Suzaku} spectral analyses, it is currently hard to tell if the \emph{Suzaku} pre-shock temperature would be constrained or significantly biased downwards even when using a correct background model.

3. Neither \emph{XMM-Newton}, nor \emph{Suzaku} detect a surface brightness discontinuity across the northern relic. In the case of \emph{XMM-Newton}, this is due to the high background level, with the ICM emission beyond the relic being consistent with zero. However, this does not appear to be the problem in the \emph{Suzaku} brightness profile. So far, all the large-scale relics with X-ray-confirmed shocks show a surface brightness jump across the relic. While not necessarily impossible (substructure along the line of sight could, in principle, complicate the brightness profile and smooth density jumps), it would be puzzling if the northern relic in CIZA J2242.8+5301 was an exception. From the paper of AK11, it is not clear whether the \emph{Suzaku} surface brightness profile was background-subtracted. If it was, the lack of a brightness discontinuity could indicate, for example, small-scale CXB fluctuations; if, let's say, the CXB is lower in the post-shock region than the assumed average, this would cause the post-shock surface brightness to be lower than that derived, and it would also bias the ICM temperatures towards smaller values.

\acknowledgements
GAO thanks Dominique Eckert and Silvano Molendi for insightful discussions during the meeting. RJwV acknowledges funding from the Royal Netherlands Academy of Arts and Sciences. MB and MH acknowledge support by the research group FOR 1254, funded by the Deutsche Forschungsgemeinschaft (DFG). AS was supported by the Einstein Postdoctoral Fellowship grant number PF9-00070, awarded by the Chandra X-ray Center, which is operated by the Smithsonian Astrophysical Observatory for NASA under contract NAS8-03060.

\end{document}